# Injection, Extraction and Matching

*M. Ferrario*
Frascati National Laboratory, National Institute for Nuclear Physics (INFN-LNF), Rome, Italy

**Abstract**
In this lecture we introduce from basic principles the main concepts of beam focusing and transport in modern accelerators using the beam envelope equation as a convenient mathematical tool. Matching conditions suitable for preserving beam quality are derived from the model for significant beam dynamics regimes. An extension of the model to the case of plasma accelerators is introduced. The understanding of similarities and differences with respect to traditional accelerators is also emphasized.

## 1    Introduction

Light sources based on high-gain free electron lasers or future high-energy linear colliders require the production, acceleration and transport up to the interaction point of low divergence, high-charge density electron bunches [1]. Many effects contribute in general to the degradation of the final beam quality, including chromatic effects, wake fields, emission of coherent radiation, and accelerator misalignments. Space charge effects and mismatch with the focusing and accelerating devices typically contribute to emittance degradation of high-charge density beams [2], hence the control of beam transport and acceleration is the leading edge for high-quality beam production.

In particular, further development of plasma-based accelerators requires careful phase space matching between plasma acceleration stages, and between plasma stages and traditional accelerator components. It represents a very critical issue and a fundamental challenge for high-quality beam production and its applications. Without proper matching, significant emittance growth may occur when the beam is propagating through different stages and components due to the large differences of transverse focusing strength. This unwanted effect is even more serious in the presence of finite energy spread.

In this paper we introduce from basic principles the main concepts of beam focusing and transport in modern accelerators using the beam envelope equation as a convenient mathematical tool. Matching conditions suitable for preserving beam quality are derived from the model for significant beam dynamics regimes. An extension of the model to the case of plasma accelerators is introduced. The understanding of similarities and differences with respect to traditional accelerators is also emphasized. A more detailed discussion of the previous topics can be found in the many classical textbooks on this subject, as listed in Refs. [3–6].

## 2    Laminar and non-laminar beams

An ideal high-charge particle beam has orbits that flow in layers that never intersect, as occurs in a laminar fluid. Such a beam is often called a laminar beam. More precisely, a laminar beam satisfies the following two conditions [6].

1. All particles at a given position have identical transverse velocities. On the contrary, the orbits of two particles that start at the same position could separate and later cross each other.
2. Assuming the beam propagates along the $z$ axis, the magnitudes of the slopes of the trajectories in the transverse directions $x$ and $y$, given by $x'(z) = dx/dz$ and $y'(z) = dy/dz$, are linearly proportional to the displacement from the $z$ axis of beam propagation.

Trajectories of interest in beam physics are always confined to the inside of small, near-axis regions, and the transverse momentum is much smaller than the longitudinal momentum, $p_{x,y} \ll p_z \approx p$. As a consequence, it is convenient in most cases to use the small angle, or *paraxial*, approximation, which allows us to write the useful approximate expressions $x' = p_x/p_z \approx p_x/p$ and $y' = p_y/p_z \approx p_x/p$.

To help understand the features and the advantages of a laminar beam propagation, the following figures compare the typical behaviour of a laminar and of a non-laminar (or thermal) beam.

Figure 1 illustrates an example of orbit evolution of a laminar mono-energetic beam with half width $x_0$ along a simple beam line with an ideal focusing element (solenoid, magnetic quadrupoles, or electrostatic transverse fields are usually adopted to this end), represented by a thin lens located at the longitudinal coordinate $z = 0$. In an ideal lens, focusing (defocusing) forces are linearly proportional to the displacement from the symmetry axis $z$ so that the lens maintains the laminar flow of the beam.

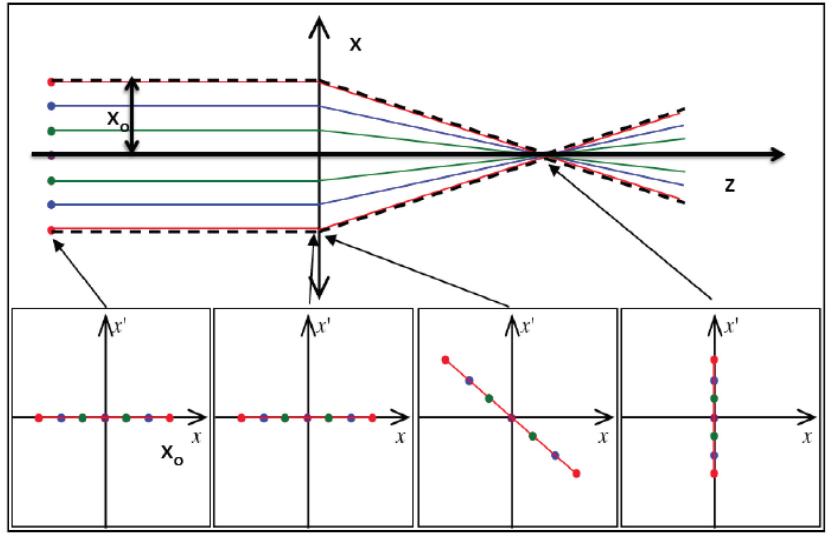

**Fig. 1:** Particle trajectories and phase space evolution of a laminar beam [7]

The beam shown in Fig. 1 starts propagating completely parallel to the symmetry axis $z$; in this particular case the particles all have zero transverse velocity. There are no orbits that cross each other in such a beam. Ignoring collisions and inner forces, like coulomb forces, such a parallel beam could propagate an infinite distance with no change in its transverse width. When the beam crosses the ideal lens it is transformed in a converging laminar beam. Because the transverse velocities after the linear lens are proportional to the displacement off-axis, particle orbits define similar triangles that converge to a single point. After passing through the singularity at the focal point, the particles follow diverging orbits. We can always transform a diverging (or converging) beam to a parallel beam by using a lens of the proper focal length, as can be seen by reversing the propagation axis of Fig. 1.

The small boxes in the lower part of the figure depict the particle distributions in the trace space $(x,x')$, equivalent to the canonical phase space $(x,p_x \approx x'p)$ when $p$ is constant, i.e. without beam acceleration. The phase space area occupied by an ideal laminar beam is a straight segment of zero thickness. As can be easily verified, the condition that the particle distribution has zero thickness proceeds from condition 1; the segment straightness is a consequence of condition 2. The distribution of a laminar beam propagating through a transport system with ideal linear focusing elements is thus a straight segment with variable slope.

Particles in a non-laminar beam have a random distribution of transverse velocities at the same location and a spread in directions, as shown in Fig. 2. Because of the disorder of a non-laminar beam, it is impossible to focus all particles from a location in the beam toward a common point. Lenses can influence only the average motion of particles. Focal spot limitations are a major concern for a wide variety of applications, from electron microscopy to free electron lasers and linear colliders. The phase space plot of a non-laminar beam is no longer a straight line: the beam, as shown in the lower boxes of Fig. 2, occupies a wider area of the phase space.

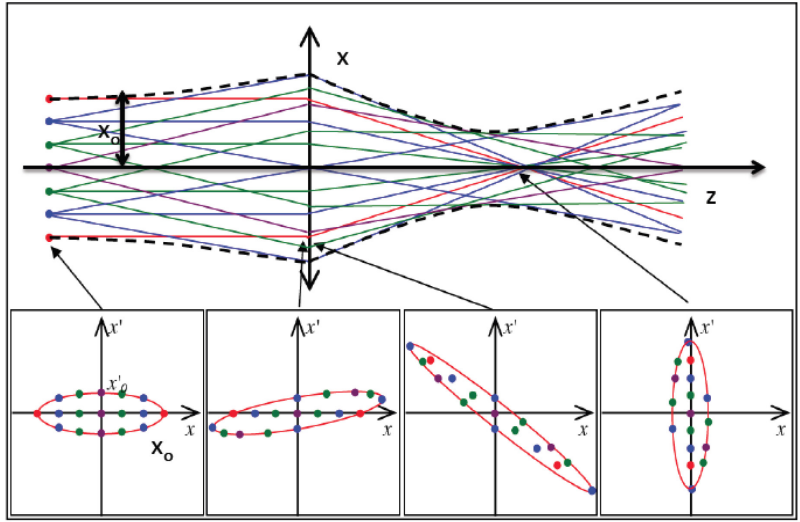

**Fig. 2:** Particle trajectories and phase space evolution of a non-laminar beam [7]

## 3    The emittance concept

The phase space surface $A$ occupied by a beam is a convenient figure of merit for designating the quality of a beam. This quantity is the emittance $\varepsilon_x$ and is usually represented by an ellipse that contains the whole particle distribution in the phase space $(x,x')$, such that $A = \pi\varepsilon_x$. An analogous definition holds for the $(y,y')$ and $(z,z')$ planes. The original choice of an elliptical shape comes from the fact that when linear focusing forces are applied to a beam, the trajectory of each particle in phase space lies on an ellipse, which may be called the trajectory ellipse. Being the area of the phase space, the emittance is measured in [mm mrad] or more often in [µm].

The ellipse equation is written as

$$\gamma_x x^2 + 2\alpha_x xx' + \beta_x x'^2 = \varepsilon_x \tag{1}$$

where $x$ and $x'$ are the particle coordinates in the phase space and the coefficients $\alpha_x(z)$, $\beta_x(z)$, $\gamma_x(z)$ are called Twiss parameters, which are related by the geometrical condition:

$$\beta_x \gamma_x - \alpha_x^2 = 1 \tag{2}$$

As shown in Fig. 3 the beam envelope boundary $X_{max}$, its derivative $(X_{max})'$ and the maximum beam divergency $X'_{max}$, i.e. the projection on the axis $x$ and $x'$ of the ellipse edges, can be expressed as a function of the ellipse parameters:

$$\begin{cases} X_{max} = \sqrt{\beta_x \varepsilon_x} \\ (X_{max})' = -\alpha\sqrt{\dfrac{\varepsilon}{\beta}} \\ X'_{max} = \sqrt{\gamma_x \varepsilon_x} \end{cases} \tag{3}$$

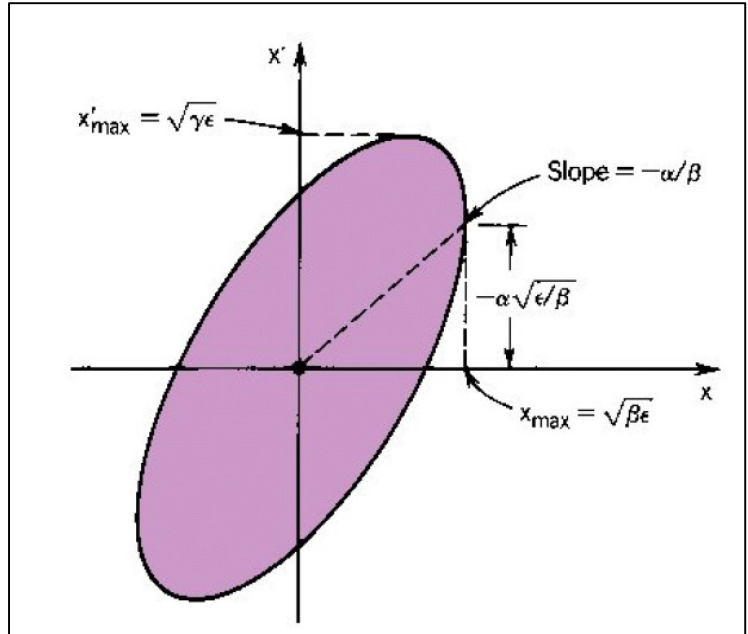

**Fig. 3:** Phase space distribution in a skewed elliptical boundary showing the relationship of Twiss parameters to the ellipse geometry [6].

According to Liouville's theorem the 6D ($x,p_x,y,p_y,z,p_z$) phase space volume occupied by a beam is constant, provided that there are no dissipative forces, no particles lost or created, and no coulomb scattering among particles. Moreover, if the forces in the three orthogonal directions are uncoupled, Liouville's theorem also holds for each reduced phase space ($x,p_x$), ($y,p_y$), ($z,p_z$) surfaces and hence emittance also remains constant in each plane [3].

Although the net phase space surface occupied by a beam is constant, nonlinear field components can stretch and distort the particle distribution in the phase space, and the beam will lose its laminar behaviour. A realistic phase space distribution is often very different from a regular ellipse, as shown in Fig. 4.

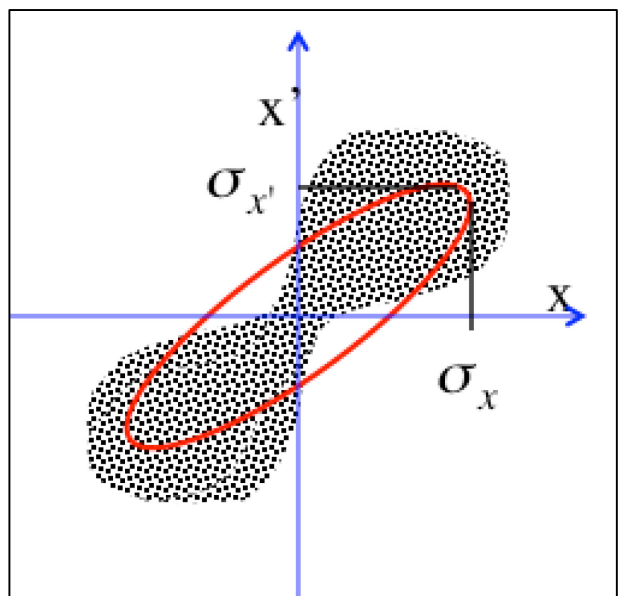

**Fig. 4:** Typical evolution of phase space distribution (black dots) under the effects of non-linear forces with the equivalent ellipse superimposed (red line).

We introduce, therefore, a definition of emittance that measures the beam quality rather than the phase space area. It is often more convenient to associate to a generic distribution function $f(x,x',z)$ in the phase space a statistical definition of emittance, the so-called *r.m.s. emittance*:

$$\gamma_x x^2 + 2\alpha_x xx' + \beta_x x'^2 = \varepsilon_{x,\text{rms}} \tag{4}$$

such that the equivalent-ellipse projections on the $x$ and $x'$ axes are equal to the r.m.s. values of the distribution, implying the following conditions:

$$\begin{cases} \sigma_x = \sqrt{\beta_x \varepsilon_{x,\text{rms}}} \\ \sigma_{x'} = \sqrt{\gamma_x \varepsilon_{x,\text{rms}}} \end{cases} \tag{5}$$

where

$$\begin{cases} \sigma_x^2(z) = \langle x^2 \rangle = \int_{-\infty}^{+\infty}\int_{-\infty}^{+\infty} x^2 f(x,x',z)\,\mathrm{d}x\mathrm{d}x' \\ \sigma_{x'}^2(z) = \langle x'^2 \rangle = \int_{-\infty}^{+\infty}\int_{-\infty}^{+\infty} x'^2 f(x,x',z)\,\mathrm{d}x\mathrm{d}x' \end{cases} \tag{6}$$

are the second moments of the distribution function $f(x,x',z)$. Another important quantity that accounts for the degree of $(x,x')$ correlations is defined as

$$\sigma_{xx'}(z) = \langle xx' \rangle = \int_{-\infty}^{+\infty}\int_{-\infty}^{+\infty} xx' f(x,x',z)\,\mathrm{d}x\mathrm{d}x'. \tag{7}$$

From Eq. (3) it holds also $\sigma'_x = \dfrac{\sigma_{xx'}}{\sigma_x} = -\alpha_x \sqrt{\dfrac{\varepsilon_{x,\text{rms}}}{\beta_x}}$, see also Eq. (16), which allows us to link the correlation moment Eq. (7) to the Twiss parameter as

$$\sigma_{xx'} = -\alpha_x \varepsilon_{x,\text{rms}} \tag{8}$$

One can easily see from Eq. (3) and Eq. (5) that $\alpha_x = -\dfrac{1}{2}\dfrac{\mathrm{d}\beta_x}{\mathrm{d}z}$ also holds.

By substituting the Twiss parameter defined by Eq. (5) and Eq. (8) into condition 2 we obtain [5]

$$\frac{\sigma_{x'}^2}{\varepsilon_{x,\text{rms}}}\frac{\sigma_x^2}{\varepsilon_{x,\text{rms}}} - \left(\frac{\sigma_{xx'}}{\varepsilon_{x,\text{rms}}}\right) = 1 \tag{9}$$

Reordering the terms in Eq. (8) we end up with the definition of *r.m.s. emittance* in terms of the second moments of the distribution:

$$\varepsilon_{\text{rms}} = \sqrt{\sigma_x^2 \sigma_{x'}^2 - \sigma_{xx'}^2} = \sqrt{\left(\langle x^2 \rangle \langle x'^2 \rangle - \langle xx' \rangle^2\right)} \tag{10}$$

where we omit, from now on, the subscribed $x$ in the emittance notation: $\varepsilon_{\text{rms}} = \varepsilon_{x,\text{rms}}$. Root mean square emittance tells us some important information about phase space distributions under the effect of linear or non-linear forces acting on the beam. Consider, for example, an idealized particle distribution in phase space that lies on some line that passes through the origin as illustrated in Fig. 5.

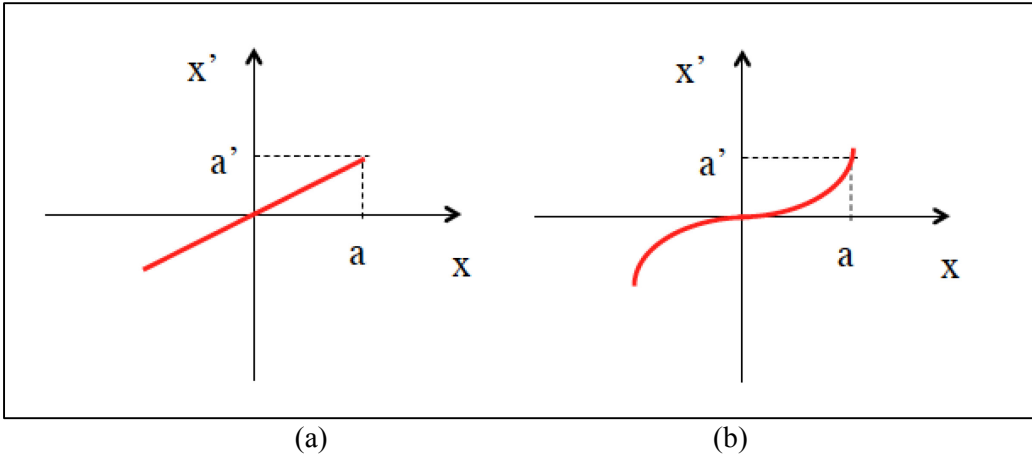

(a)             (b)
**Fig. 5:** Phase space distributions under the effect of (a) linear or (b) non-linear forces acting on the beam

Assuming a generic correlation of the type $x' = Cx^n$ and computing the r.m.s. emittance according to Eq. (10) we have

$$\varepsilon_{rms}^2 = C\sqrt{\langle x^2\rangle\langle x^{2n}\rangle - \langle x^{n+1}\rangle^2} \quad \begin{cases} n=1 \Rightarrow \varepsilon_{rms} = 0 \\ n>1 \Rightarrow \varepsilon_{rms} \neq 0 \end{cases}. \quad (11)$$

When $n = 1$ the line is straight and the r.m.s. emittance is $\varepsilon_{rms} = 0$. When $n > 1$ the relationship is nonlinear, the line in phase space is curved, and the r.m.s. emittance is in general not zero. Both distributions have zero area. Therefore, we conclude that even when the phase space area is zero, if the distribution is lying on a curved line its r.m.s. emittance is not zero. The r.m.s. emittance depends not only on the area occupied by the beam in phase space, but also on distortions produced by non-linear forces.

If the beam is subject to acceleration it is more convenient to use the r.m.s. normalized emittance, for which the transverse momentum $p_x = p_z x' = m_o c \beta \gamma x'$ is used instead of the divergence:

$$\varepsilon_{n,rms} = \frac{1}{m_0 c}\sqrt{\sigma_x^2 \sigma_{p_x}^2 - \sigma_{xp_x}^2} = \frac{1}{m_0 c}\sqrt{(\langle x^2\rangle\langle p_x^2\rangle - \langle xp_x\rangle^2)} = \sqrt{(\langle x^2\rangle\langle(\beta\gamma x')^2\rangle - \langle x\beta\gamma x'\rangle^2)}. \quad (12)$$

The reason for introducing a normalized emittance is that the divergences of the particles $x' = p_x/p$ are reduced during acceleration as $p$ increases. Thus, acceleration reduces the un-normalized emittance, but does not affect the normalized emittance. Assuming a small energy spread within the beam, the normalized and un-normalized emittances can be related by the approximated relation $\langle\beta\gamma\rangle\varepsilon_{rms}$. This approximation, which is often used in conventional accelerators, may be strongly misleading when adopted for describing beams with significant energy spread, like those presently produced by plasma accelerators. A more careful analysis is reported below [8].

When the correlations between the energy and transverse positions are negligible (as in a drift without collective effects), Eq. (12) can be written as

$$\varepsilon_{n,rms}^2 = \langle\beta^2\gamma^2\rangle\langle x^2\rangle\langle x'^2\rangle - \langle\beta\gamma\rangle^2\langle xx'\rangle^2. \quad (13)$$

Consider now the definition of relative energy spread $\sigma_\gamma^2 = \dfrac{\langle\beta^2\gamma^2\rangle - \langle\beta\gamma\rangle^2}{\langle\beta\gamma\rangle^2}$, which can be inserted into Eq. (13) to give

$$\varepsilon_{n,rms}^2 = \langle\beta^2\gamma^2\rangle\sigma_\gamma^2\langle x^2\rangle\langle x'^2\rangle + \langle\beta\gamma\rangle^2\left(\langle x^2\rangle\langle x'^2\rangle - \langle xx'\rangle^2\right). \quad (14)$$

Assuming relativistic particles ($\beta = 1$) we get

$$\varepsilon_{n,\text{rms}}^2 = \langle \gamma^2 \rangle \left( \sigma_\gamma^2 \sigma_x^2 \sigma_{x'}^2 + \varepsilon_{\text{rms}}^2 \right). \tag{15}$$

If the first term in the parentheses is negligible, we find the conventional approximation of normalized emittance, as $\langle \gamma \rangle \varepsilon_{\text{rms}}$. For a conventional accelerator this might generally be the case. Considering, for example, beam parameters for the SPARC_LAB photoinjector [9], at 5 MeV the ratio between the first and the second term is ~$10^{-3}$; while at 150 MeV it is ~$10^{-5}$. On the other hand, using typical beam parameters at the plasma–vacuum interface, the first term is of the same order of magnitude as for conventional accelerators at low energies; however, due to the rapid increase of the bunch size outside the plasma ($\sigma_{x'} \sim$ mrad) and the large energy spread ($\sigma_\gamma > 1\%$), it becomes predominant compared to the second term after a drift of a few millimetres. *Therefore the use of approximated formulas when measuring the normalized emittance of plasma accelerated particle beams is very inappropriate* [10].

## 4 The root mean square envelope equation

We are now interested in following the evolution of particle distribution during beam transport and acceleration. One can take profit of the first collective variable defined in Eq. (6), the second moment of the distribution termed r.m.s. beam envelope, to derive a differential equation suitable for describing r.m.s. beam envelope dynamics [11]. To this end let us compute the first and second derivative of $\sigma_x$ [4]:

$$\begin{aligned}\frac{d\sigma_x}{dz} &= \frac{d}{dz}\sqrt{\langle x^2 \rangle} = \frac{1}{2\sigma_x}\frac{d}{dz}\langle x^2 \rangle = \frac{1}{2\sigma_x}2\langle xx' \rangle = \frac{\sigma_{xx'}}{\sigma_x} \\ \frac{d^2\sigma_x}{dz^2} &= \frac{d}{dz}\frac{\sigma_{xx'}}{\sigma_x} = \frac{1}{\sigma_x}\frac{d\sigma_{xx'}}{dz} - \frac{\sigma_{xx'}^2}{\sigma_x^3} = \frac{1}{\sigma_x}\left(\langle x'^2 \rangle + \langle xx'' \rangle\right) - \frac{\sigma_{xx'}^2}{\sigma_x^3} = \frac{\sigma_{x'}^2 + \langle xx'' \rangle}{\sigma_x} - \frac{\sigma_{xx'}^2}{\sigma_x^3} \end{aligned}. \tag{16}$$

Rearranging the second derivative Eq. (16) we obtain a second-order nonlinear differential equation for the beam envelope evolution,

$$\sigma_x'' = \frac{\sigma_x^2 \sigma_{x'}^2 - \sigma_{xx'}^2}{\sigma_x^3} + \frac{\langle xx'' \rangle}{\sigma_x} \tag{17}$$

or, in a more convenient form using the r.m.s. emittance definition Eq. (10),

$$\sigma_x'' - \frac{1}{\sigma_x}\langle xx'' \rangle = \frac{\varepsilon_{\text{rms}}^2}{\sigma_x^3}. \tag{18}$$

In Eq. (18) the emittance term can be interpreted physically as an outward pressure on the beam envelope produced by the r.m.s. spread in trajectory angle, which is parameterized by the r.m.s. emittance.

Let's now consider, for example, the simple case with $\langle xx'' \rangle = 0$, describing a beam drifting in free space. The envelope equation reduces to

$$\sigma_x^3 \sigma_x'' = \varepsilon_{\text{rms}}^2. \tag{19}$$

With initial conditions $\sigma_0$, $\sigma'_0$ at $z_0$, depending on the upstream transport channel, Eq. (19) has a hyperbolic solution:

$$\sigma(z) = \sqrt{\left(\sigma_0 + \sigma'_0(z - z_0)\right)^2 + \frac{\varepsilon_{\text{rms}}^2}{\sigma_0^2}(z - z_0)^2}. \tag{20}$$

Considering the case of a beam at waist ($\langle xx' \rangle = 0$) with $\sigma'_0 = 0$, using Eq. (5), the solution Eq. (20) is often written in terms of the $\beta$ function as

$$\sigma(z) = \sigma_0 \sqrt{1 + \left(\frac{z-z_0}{\beta_w}\right)^2} \qquad (21)$$

This relation indicates that without any external focusing element the beam envelope increases from the beam waist by a factor $\sqrt{2}$ with a characteristic length $\beta_w = \frac{\sigma_0^2}{\epsilon_{rms}}$ as shown in Fig. 6.

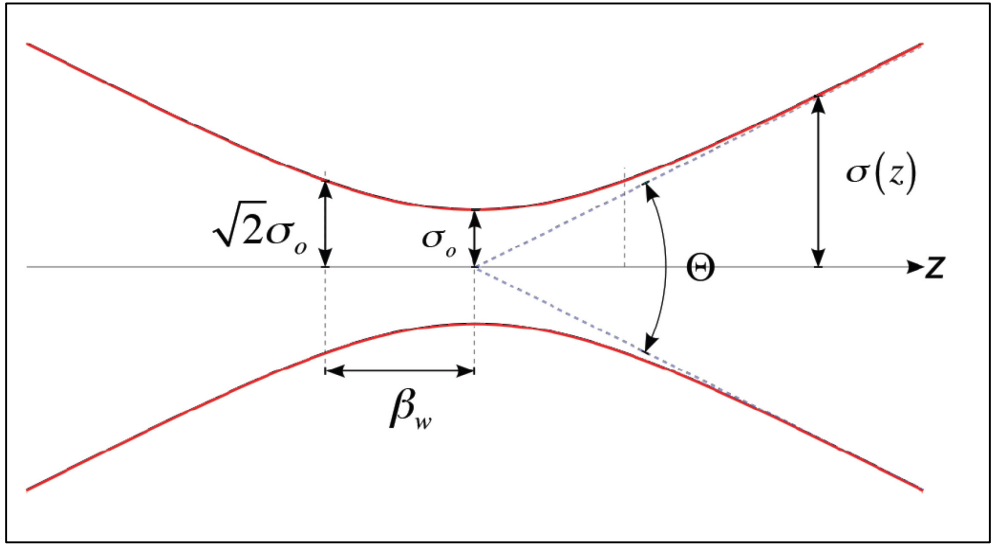

**Fig. 6:** Schematic representation of the beam envelope behaviour near the beam waist

At waist the relation $\varepsilon^2_{rms} = \sigma^2_{0,x}\sigma^2_{0,x'}$ also holds, which can be inserted into Eq. (20) to get $\sigma^2_x(z) = \sigma^2_{0x'}(z-z_0)^2$. Under this condition Eq. (15) can be written as follows:

$$\varepsilon^2_{n,rms}(z) = \langle \gamma^2 \rangle \left( \sigma_\gamma^2 \sigma_{x'}^4 (z-z_0)^2 + \varepsilon^2_{rms} \right)$$

*showing that beams with large energy spread and divergence undergo a significant normalized emittance growth even in a drift of length $(z - z_0)$ [8, 12].*

Notice also that the solution Eq. (21) is exactly analogous to that of a Gaussian light beam for which the beam width $w = 2\sigma_{ph}$ increases away from its minimum value at the waist $w_0$ with characteristic length $Z_R = \frac{\pi w_0^2}{\lambda}$ (Rayleigh length) [4]. This analogy suggests that we can identify an effective emittance of a photon beam as $\varepsilon_{ph} = \frac{\lambda}{4\pi}$.

For the effective transport of a beam with finite emittance it is mandatory to make use of some external force providing beam confinement in the transport or accelerating line. The term $\langle xx'' \rangle$ accounts for external forces when we know $x''$ given by the single particle equation of motion:

$$\frac{dp_x}{dt} = F_x. \qquad (22)$$

Under the paraxial approximation $p_x \ll p = \beta\gamma mc$ the transverse momentum $p_x$ can be written as $p_x = px' = \beta\gamma m_0 cx'$, so that

$$\frac{\mathrm{d}p_x}{\mathrm{d}t} = \frac{\mathrm{d}}{\mathrm{d}t}(px') = \beta c \frac{\mathrm{d}}{\mathrm{d}z}(px') = F_x \tag{23}$$

and the transverse acceleration results to be:

$$x'' = -\frac{p'}{p}x' + \frac{F_x}{\beta c p}. \tag{24}$$

It follows that

$$\langle xx'' \rangle = -\frac{p'}{p}\langle xx' \rangle + \frac{\langle xF_x \rangle}{\beta c p} = \frac{p'}{p}\sigma_{xx'} + \frac{\langle xF_x \rangle}{\beta c p}. \tag{25}$$

Inserting Eq. (25) into Eq. (18) and recalling Eq. (16), $\sigma'_x = \frac{\sigma_{xx'}}{\sigma_x}$, the complete r.m.s. envelope equation is

$$\sigma''_x + \frac{p'}{p}\sigma'_x - \frac{1}{\sigma_x}\frac{\langle xF_x \rangle}{\beta c p} = \frac{\varepsilon^2_{n,\mathrm{rms}}}{\gamma^2 \sigma^3_x} \tag{26}$$

where we have included the normalized emittance $\varepsilon_{n,\mathrm{rms}} = \gamma\varepsilon_{\mathrm{rms}}$. Notice that the effect of longitudinal accelerations appears in the r.m.s. envelope equation as an oscillation damping term, called 'adiabatic damping', proportional to $p'/p$. The term $\langle xF_x \rangle$ represents the moment of any external transverse force acting on the beam, such as that produced by a focusing magnetic channel.

## 5 External forces

Let's now consider the case of an external linear force acting on the beam in the form $F_x = \pm kx$. It can be focusing or defocusing according to the sign. The moment of the force results as

$$\langle xF_x \rangle = \pm k \langle x^2 \rangle = \pm k\sigma^2_x \tag{27}$$

and the envelope equation becomes

$$\sigma''_x + \frac{\gamma'}{\gamma}\sigma'_x \mp k^2_{\mathrm{ext}}\sigma_x = \frac{\varepsilon^2_{n,\mathrm{rms}}}{\gamma^2 \sigma^3_x} \tag{28}$$

where we have explicitly used the momentum definition $p = \gamma mc$ for a relativistic particle with $\beta \approx 1$ and defined the wavenumber $k^2_{\mathrm{ext}} = \frac{k}{\gamma m_0 c^2}$.

Typical focusing elements are quadrupoles and solenoids [3]. The magnetic quadrupole field is given in Cartesian coordinates by

$$\begin{cases} B_x = B_0 \dfrac{y}{\mathrm{d}} = B'_0 y \\ B_y = B_0 \dfrac{x}{\mathrm{d}} = B'_0 x \end{cases} \tag{29}$$

where $d$ is the pole distance and $B'_0$ the field gradient. The force acting on the beam is $\vec{F}_\perp = qv_z B'_0 (y\hat{j} - x\hat{i})$ that, when $B_0$ is positive, is focusing in the $x$ direction and defocusing in $y$. The focusing strength is $k_{quad} = \dfrac{qB'_0}{\gamma m_0 c} = k_{ext}^2$.

In a solenoid the focusing strength is given by $k_{sol} = \left(\dfrac{qB_0}{2\gamma m_0 c}\right)^2 = k_{ext}^2$. Notice that the solenoid is always focusing in both directions, an important properties when the cylindrical symmetry of the beam must be preserved. On the other hand, being a second-order quantity in $\gamma$ it is more effective at low energy.

It is interesting to consider the case of a uniform focusing channel without acceleration described by the r.m.s. envelope equation

$$\sigma''_x + k_{ext}^2 \sigma_x = \dfrac{\varepsilon_{rms}^2}{\sigma_x^3}. \tag{30}$$

By substituting $\sigma_x = \sqrt{\beta_x \varepsilon_{rms}}$ in Eq. (30) one obtains an equation for the 'betatron function' $\beta_x(z)$ that is independent of the emittance term:

$$\beta''_x + 2k_{ext}^2 \beta_x = \dfrac{2}{\beta_x} + \dfrac{\beta'^2_x}{2\beta_x}. \tag{31}$$

Equation (31) contains just the transport channel focusing strength and, being independent of the beam parameters, suggests that the meaning of the betatron function is to describe the transport line characteristic by itself. The betatron function reflects exterior forces from focusing magnets, and is highly dependent on the particular arrangement of the quadrupole magnets. The equilibrium, or matched, solution of Eq. (31) is given by $\beta_{eq} = \dfrac{1}{k_{ext}} = \dfrac{\lambda_\beta}{2\pi}$, as can be easily verified. This result shows that the matched $\beta_x$ function is simply the inverse of the focusing wave number, or equivalently is proportional to the 'betatron wavelength' $\lambda_\beta$.

# 6  Space charge forces

Another important force acting on the beam is the one produced by the beam itself due to the internal coulomb forces. The net effect of the coulomb interaction in a multi-particle system can be classified into two regimes [3]:

– *collisional regime*, dominated by binary collisions caused by close particle encounters;
– *collective regime* or *space charge regime*, dominated by the self-field produced by the particles' distribution that varies appreciably only over large distances compared to the average separation of the particles.

A measure for the relative importance of collisional versus collective effects in a beam with particle density $n$ is the relativistic *Debye length*,

$$\lambda_D = \sqrt{\dfrac{\varepsilon_0 \gamma^2 k_B T_b}{e^2 n}} \tag{32}$$

where the transverse beam temperature $T_b$ is defined as $k_B T_b = \gamma m_0 \langle v_\perp^2 \rangle$, and $k_B$ is the Boltzmann constant. As long as the Debye length remains small compared to the particle bunch transverse size the beam is in the space-charge dominated regime and is not sensitive to binary collisions. Smooth functions for the charge and field distributions can be used in this case, and the space charge force can be treated like an external applied force. The space charge field can be separated into linear and nonlinear terms as a function of displacement from the beam axis. The linear space charge term defocuses the beam and leads to an increase in beam size. The nonlinear space charge terms also increase the r.m.s. emittance by distorting the phase-space distribution. Under the paraxial approximation of particle motion we can consider the linear component alone. We shall see below that the linear component of the space charge field can also induce emittance growth when correlation along the bunch are taken into account.

For a bunched beam of uniform charge distribution in a cylinder of radius $R$ and length $L$, carrying a current $\hat{I}$ and moving with longitudinal velocity $v_z = \beta c$, the linear component of the longitudinal and transverse space charge field are given approximately by [13]

$$E_z(\zeta) = \frac{\hat{I}L}{2\pi\varepsilon_0 R^2 \beta c} h(\zeta), \tag{33}$$

$$E_r(r,\zeta) = \frac{\hat{I}r}{2\pi\varepsilon_0 R^2 \beta c} g(\zeta). \tag{34}$$

The field form factor is described by the functions

$$h(\zeta) = \sqrt{A+(1-\zeta)^2} - \sqrt{A+\zeta^2 + (2\zeta - 1)}, \tag{35}$$

$$g(\zeta) = \frac{(1-\zeta)}{2\sqrt{A^2+(1-\zeta)^2}} + \frac{\zeta}{2\sqrt{A^2+\zeta}} \tag{36}$$

where $\zeta = z/L$ is the normalized longitudinal coordinate along the bunch and $A = R/\gamma L$ is the beam aspect ratio. The field form factors account for the variation of the fields along the bunch. As $\gamma$ increases, $g(\zeta) \to 1$ and $h(\zeta) \to 0$, thus showing that space charge fields mainly affect transverse beam dynamics. It shows also that an energy increase corresponds to a bunch lengthening in the moving frame $L' = \gamma L$, leading to a vanishing longitudinal field component, as in the case of a continuous beam in the laboratory frame.

To evaluate the force acting on the beam one must also account for the azimuthal magnetic field associated with the beam current that in cylindrical symmetry is given by $B_\vartheta = \frac{\beta}{c} E_r$. Thus, the Lorentz force acting on each single particle is given by

$$F_r = e(E_r - \beta c B_\vartheta) = e(1-\beta^2)E_r = \frac{eE_r}{\gamma^2}. \tag{37}$$

The attractive magnetic force, which becomes significant at high velocities, tends to compensate for the repulsive electric force. Therefore space charge defocusing is primarily a non-relativistic effect and decreases as $\gamma^{-2}$.

In order to include space charge forces in the envelope equation let us start writing the space charge forces produced by the previous fields in Cartesian coordinates:

$$F_x = \frac{e\hat{I}x}{2\pi\gamma^2\varepsilon_0 \sigma_x^2 \beta c} g(\zeta). \tag{38}$$

Then, computing the moment of the force we need

$$x'' = \frac{F_x}{\beta c p} = \frac{eIx}{2\pi\varepsilon_0 \gamma^3 m_0 \beta^3 c^3 \sigma_x^2} = \frac{k_{sc}(\zeta)}{(\beta\gamma)^3 \sigma_x^2} \quad (39)$$

where we have introduced the generalized beam perveance

$$k_{sc}(\zeta) = \frac{2\hat{I}}{I_A} g(\zeta) \quad (40)$$

normalized to the Alfven current $I_A = 4\pi\varepsilon_0 m_0 c^3/e = 17$ kA for electrons. Notice that in this case the perveance in Eq. (40) explicitly depends on the slice coordinate $\zeta$. We can now calculate the term that enters in the envelope equation for a relativistic beam,

$$\langle xx'' \rangle = \frac{k_{sc}}{\gamma^3 \sigma_x^2} \langle x^2 \rangle = \frac{k_{sc}}{\gamma^3}, \quad (41)$$

leading to the complete envelope equation

$$\sigma_x'' + \frac{\gamma'}{\gamma}\sigma_x' + k_{ext}^2 \sigma_x = \frac{\varepsilon_{n,rms}^2}{\gamma^2 \sigma_x^3} + \frac{k_{sc}}{\gamma^3 \sigma_x}. \quad (42)$$

From the envelope equation Eq. (42) we can identify two regimes of beam propagation: *space-charge dominated* and *emittance dominated*. A beam is space-charge dominated as long as the space charge collective forces are largely dominant over the emittance pressure. In this regime the linear component of the space-charge force produces a quasi-laminar propagation of the beam, as one can see by integrating one time Eq. (39) under the paraxial ray approximation $x' \ll 1$. A measure of the relative importance of space-charge effects versus emittance pressure is given by the *laminarity parameter*, defined as the ratio between the space-charge term and the emittance term:

$$\rho = \frac{\hat{I}}{2I_A \gamma} \frac{\sigma^2}{\varepsilon_n^2}. \quad (43)$$

When $\rho$ greatly exceeds unity, the beam behaves like a laminar flow (all beam particles move on trajectories that do not cross), and transport and acceleration require a careful tuning of focusing and accelerating elements in order to keep laminarity. Correlated emittance growth is typical in this regime, which can be made reversible if proper beam matching conditions are fulfilled, as discussed below. When $\rho < 1$ the beam is emittance dominated (thermal regime) and the space charge effects can be neglected. The transition to the thermal regime occurs when $\rho \approx 1$ corresponding to the transition energy

$$\gamma_{tr} = \frac{\hat{I}}{2I_A} \frac{\sigma^2}{\varepsilon_n^2}. \quad (44)$$

For example a beam with $\hat{I} = 100$ A $\varepsilon_n = 1$ μm and $\sigma = 300$ μm is leaving the space charge dominated regime and is entering the thermal regime at the transition energy of 131 MeV. From this example one may conclude that the space charge dominated regime is typical of low energy beams. Actually, for applications like linac-driven free electron lasers, peak current exceeding kilo amperes are required. Space charge effects may recur if bunch compressors are active at higher energies and a new energy threshold with higher $\hat{I}$ has to be considered.

## 7 Correlated emittance oscillations

When longitudinal correlations within the bunch are important, like that induced by space charge effects, beam envelope evolution is generally dependent also on the bunch coordinate $\zeta$. In this case the bunch should be considered as an ensemble of *n* longitudinal slices of envelope $\sigma_s(z,\zeta)$, whose evolution can be computed from *n* slice envelope equations equivalent to Eq. (42) provided that the bunch parameters refer to each single slice: $\gamma_s$, $\gamma'_s$, $k_{sc,s} = k_{sc}g(\zeta)$. Correlations within the bunch may cause emittance oscillations that can be evaluated, once an analytical or numerical solution [13] of the slice envelope equation is known, by using the following correlated emittance definition:

$$\varepsilon_{rms,cor} = \sqrt{\langle\sigma_s^2\rangle\langle\sigma_s'^2\rangle - \langle\sigma_s\sigma_s'\rangle^2} \qquad (45)$$

where the average is performed over the entire slice ensemble. In the simplest case of a two-slices model the previous definition reduces to

$$\varepsilon_{rms,cor} = |\sigma_1\sigma_2' - \sigma_2\sigma_1'| \, , \qquad (46)$$

which represents a simple and useful formula for an estimation of the emittance scaling [14].

The total normalized r.m.s. emittance is the given by the superposition of the correlated and uncorrelated terms as

$$\varepsilon_{rms,cor} = \langle\gamma\rangle\sqrt{\varepsilon_{rms}^2 + \varepsilon_{rms,cor}^2} \, . \qquad (47)$$

An interesting example to consider here, showing the consequences of non-perfect beam matching, is the propagation of a beam in the space-charge dominated regime nearly matched to an external focusing channel ($k_{ext} = k_{sol}$), as illustrated in Fig. 7. To simplify our computations we can neglect acceleration, as in the case of a simple beam transport line. The envelope equation for each slice, indicated as $\sigma_s$, reduces to

$$\sigma_s'' + k_{ext}^2\sigma_s = \frac{k_{sc,s}}{\gamma^3\sigma_s} \, . \qquad (48)$$

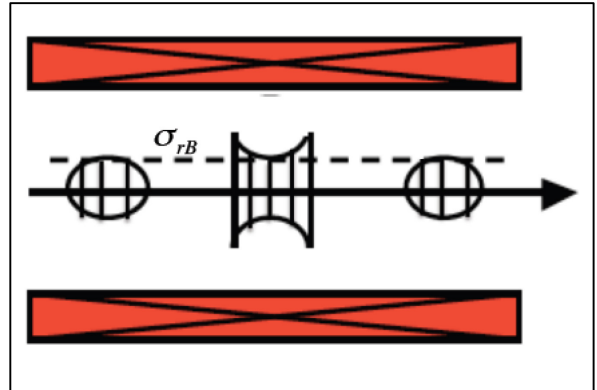

**Fig. 7:** Schematic representation of a nearly matched beam in a long solenoid. The dashed line represent the reference slice envelope fully matched to the Brillouin flow condition. The other slice envelopes are oscillating around the equilibrium solution.

A stationary solution, called the *Brillouin flow*, is given by

$$\sigma_{s,B} = \frac{1}{k_{ext}^2}\sqrt{\frac{\hat{I}g(\zeta)}{2\gamma^3 I_A}} \qquad (49)$$

where the local dependence of the current $\hat{I}_s = \hat{I}g(\zeta)$ within the bunch has been explicitly indicated. This solution represent the matching conditions for which the external focusing completely balances the internal space-charge force. Unfortunately, since $k_{ext}$ has a slice-independent constant value, the Brillouin matching condition cannot be achieved at the same time for all of the bunch slices. Assuming that there is a reference slice perfectly matched with an envelope $\sigma_{r,B}$, the matching condition for the other slices can be written as

$$\sigma_{sB} = \sigma_{rB} + \frac{\sigma_{rB}}{2}\left(\frac{\delta I_s}{\hat{I}}\right) \tag{50}$$

with respect to the reference slice. Considering a small perturbation $\delta_s$ from the equilibrium in the form

$$\sigma_s = \sigma_{s,B} + \delta_s \tag{51}$$

and substituting into Eq. (48) we can obtain a linearized equation for the slice offset

$$\delta_s'' + 2k_{ext}^2 \delta_s = 0 \tag{52}$$

which has a solution given by

$$\delta_s = \delta_0 \cos\left(\sqrt{2}k_{ext}z\right) \tag{53}$$

where $\delta_0 = \sigma - \sigma_{sB}$ is the amplitude of the initial slice mismatch, which we assume for convenience is the same for all slices. Inserting Eq. (53) into Eq. (51) we get the perturbed solution:

$$\sigma_s = \sigma_{s,B} + \delta_0 \cos\left(\sqrt{2}k_{ext}z\right). \tag{54}$$

Equation (54) shows that slice envelopes oscillate together around the equilibrium solution with the same frequency for all slices ($\sqrt{2}k_{ext}$, often called the plasma frequency) dependent only on the external focusing forces. This solution represents a collective behaviour of the bunch similar to that of the electrons subject to the restoring force of ions in a plasma. Using the two-slices model and Eq. (54) the emittance evolution Eq. (46) results in

$$\varepsilon_{rms,cor} = \frac{1}{4}k_{sol}\sigma_{rB}\left|\frac{\Delta I}{\hat{I}}\delta_0 \sin\left(\sqrt{2}k_{ext}z\right)\right| \tag{55}$$

where $\Delta I = \hat{I}_1 - \hat{I}_2$. Notice that, in this simple case, envelope oscillations of the mismatched slices induce correlated emittance oscillations that periodically go back to zero, showing the reversible nature of the correlated emittance growth. It is, in fact, the coupling between transverse and longitudinal motion induced by the space-charge fields that allows reversibility. With a proper tuning of the transport line length or of the focusing field one can compensate for the transverse emittance growth at the expenses of the longitudinal emittance.

At first it may seem surprising that a beam with a single charge species can exhibit plasma oscillations, which are characteristic of plasmas composed of two-charge species. But the effect of the external focusing force can play the role of the other charge species, providing the necessary restoring force that is the cause of such collective oscillations, as shown in Fig. 8. The beam can actually be considered as a single component, relativistic, cold plasma.

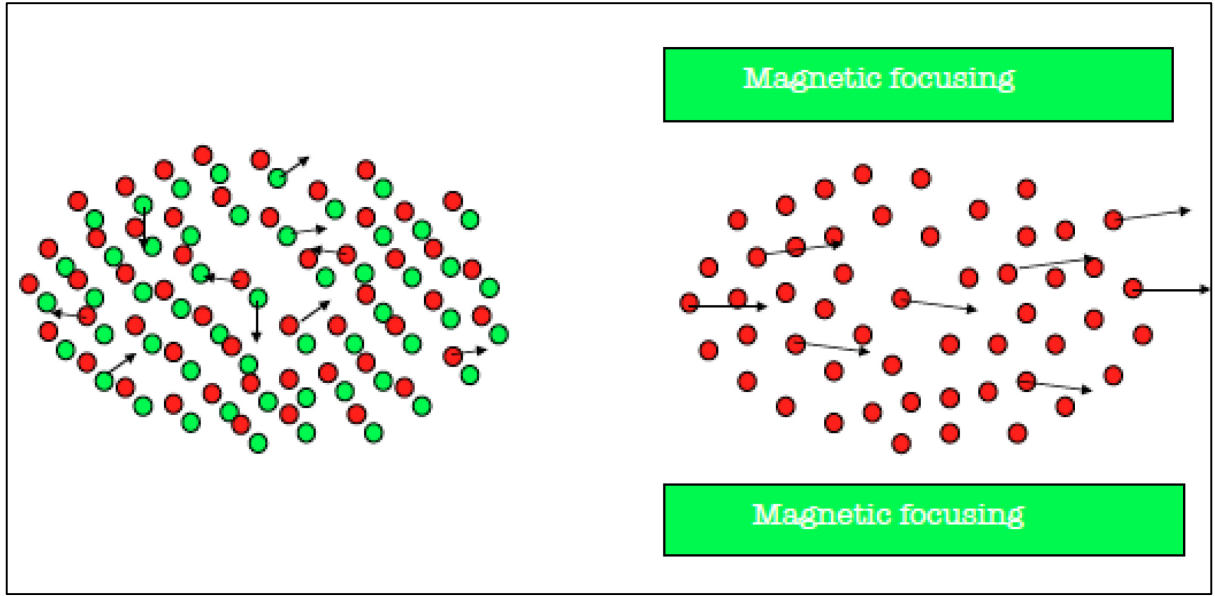

**Fig. 8:** The restoring force produced by the ions (green dots) in a plasma may cause electron (red dots) oscillations around the equilibrium distribution. In a similar way the restoring force produced by a magnetic field may cause beam envelope oscillations around the matched envelope equilibrium.

It is important to bear in mind that beams in linacs are also different from plasmas in some important respects [5]. One is that beam transit time through a linac is too short for the beam to reach thermal equilibrium. Also, unlike a plasma, the Debye length of the beam may be larger than, or comparable to, the beam radius, so shielding effects may be incomplete.

## 8  Matching conditions in a plasma accelerator

The concepts developed for the beam transport in the previous sections can be now applied to the case of a plasma accelerator [15], giving important information about the critical topic of beam–plasma matching conditions. To this end we introduce a simplified model for the plasma and for the resulting fields acting on the beam in order to be able to write an envelope equation for the accelerated beam.

In this section we are interested in the case of the external injection of particles in a plasma wave, in the so-called 'bubble' regime, that could be excited by a short, intense laser pulse [15, 16] or by a driving electron beam [17, 18] with beam density $n_b$ larger than the plasma density $n_0$, $n_b > n_0$. A very simplified model for the plasma behind the driving pulse is illustrated in Fig. 9. We will consider a spherical, uniform ion distribution, as indicated by a dashed circle, with particle density $n_0$. This model is justified by the fact that, in this regime, the fields are linear in longitudinal and transverse directions, at least in the region of interest for particle acceleration, as that produced by a uniform ion distribution within a sphere of radius $R_{\text{sphere}} \approx \lambda_p / 2$ where $\lambda_p = 2\pi c \sqrt{\varepsilon_0 m / n_0 e^2}$ is the plasma wavelength. A more detailed treatment [19] shows that the correct scaling is $R_{\text{sphere}} = 2\sqrt{(n_b / n_0)}\sigma_r$, where $\sigma_r$ is the driving beam r.m.s. radius, that for a uniform cylindrical driving bunch gives $R_{\text{sphere}} = \sqrt{\dfrac{4eI}{\pi^3 mc^3 \varepsilon_0}} \dfrac{\lambda_p}{2}$.

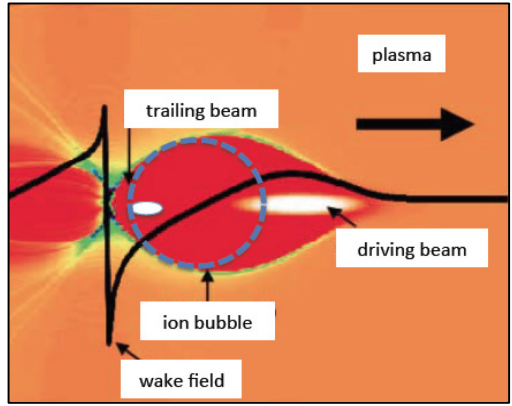

**Fig. 9:** Schematic representation of the longitudinal wake field (black line) and ion distribution (red area) behind a driving laser or particle beam [16].

The field produced by the ions and experienced by a witness electron beam is purely electrostatic, being the ions at rest in the laboratory frame in the timescale of interest, and is simply given by

$$E_r = \frac{en_0}{3\varepsilon_0} r, \qquad (56)$$

i.e. it has a radial symmetry (other authors, see for example Ref. [17], consider a uniform charged cylindrical ion column producing a transverse field of the form $E_r = \frac{en_0}{2\varepsilon_0} r$). The ion sphere is 'virtually' moving along $z$ with the speed $\beta_d$ of the driving pulse due to the plasma electron collective oscillation, even if the source of the field remains at rest in the laboratory frame. There are also magnetic fields produced by the plasma electron displacement but, as shown in Ref. [20], the net effect on a relativistic beam is negligible.

The accelerating component of the field is linearly increasing from the moving sphere centre $z_c = \beta_d ct$:

$$E_z(\zeta) = \frac{en_0}{3\varepsilon_0} \zeta \qquad (57)$$

where $\zeta = z - z_c$, and has a maximum on the sphere edge at $\zeta = \lambda_p / 2$. The corresponding energy gained by a witness electron is given by $\gamma = \gamma_0 + \alpha L_{acc}$ where $L_{acc}$ is the accelerating length in the plasma and $\alpha(\zeta) = \frac{eE_z(\zeta)}{mc^2} = \frac{1}{3}\left(\frac{2\pi c}{\lambda_p}\right)^2 \zeta$ is the normalized accelerating gradient. The energy spread accumulated by a bunch of finite r.m.s. length $\sigma_z$ is given by $\frac{\delta\gamma}{\gamma} = \frac{\delta\alpha L_{acc}}{\gamma_0 + \alpha L_{acc}} \approx \frac{\delta\alpha}{\alpha} = \frac{\sigma_z}{\lambda_p}$, showing that ultra-short electron bunches are required to keep energy spread below 1%. In this simplified model, beam loading effects are not considered, nor beam slippage with respect to the driving pulse.

The transverse (focusing) field

$$E_x = \frac{en_0}{3\varepsilon_0} x \qquad (58)$$

at a distance $x$ off the propagation axis is independent of $\zeta$ so that correlated emittance growth is not typically induced by the ion focusing field.

In Fig. 10 are shown the plasma wavelength and the longitudinal and transverse fields experienced by a test particle located at $x = 1$ μm and $\zeta = \lambda_p / 4$ versus typical plasma densities, according to Eqs. (54, 55).

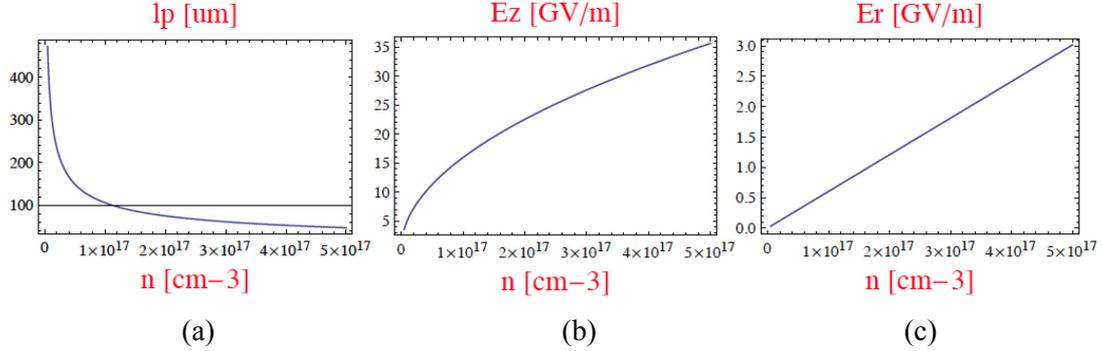

(a)            (b)            (c)

**Fig. 10:** (a) Plasma wavelength, (b) longitudinal and (c) transverse fields versus typical plasma densities experienced by a test particle located at $x = 1$ μm and $\zeta = \lambda_p / 4$.

As discussed in the previous sections the transverse beam dynamics can be conveniently described by means of a proper envelope equation. To this end let us consider the single particle equation of motion:

$$x'' = \frac{F_x}{\beta c p} = \frac{e^2 n_0}{3\varepsilon_0 \gamma m c^2} x = \frac{k_p^2}{3\gamma} x \qquad (59)$$

where $k_p = \sqrt{n e^2 / \varepsilon_0 m c^2}$ is the plasma wave number. The moment of the force acting on the beam particles is given by

$$\langle x x'' \rangle = \frac{k_p^2}{3\gamma} \langle x^2 \rangle = \frac{k_p^2}{3\gamma} \sigma_x^2 . \qquad (60)$$

Inserting into the envelope equation we obtain

$$\sigma_x'' + \frac{\gamma'}{\gamma} \sigma_x' + \frac{k_p^2}{3\gamma} \sigma_x = \frac{\varepsilon_n^2}{\gamma^2 \sigma_x^3} + \frac{k_{sc}^0}{\gamma^3 \sigma_x}. \qquad (61)$$

An equilibrium solution of the previous equation has not yet been found, nevertheless some simplification is still possible and an approximated matching condition exists. As one can see, there are two focusing terms, the adiabatic damping and the ion focusing, and two defocusing terms, the emittance pressure and the space-charge effects. To compare the relative importance of the first two terms it is more convenient to rewrite the previous equation with the new variable $\tilde{\sigma}_x = \sqrt{\gamma} \sigma_x$ leading to the equation

$$\tilde{\sigma}_x'' + \left( \left( \frac{\gamma'}{2\gamma} \right)^2 + \frac{k_p^2}{3\gamma} \right) \tilde{\sigma}_x = \frac{\varepsilon_n^2}{\tilde{\sigma}_x^3} + \frac{k_o^{sc}}{\gamma^2 \tilde{\sigma}_x}. \qquad (62)$$

The beam is space charge dominated, as already discussed in Section 6, when

$$\rho = \frac{k_o^{sc} \tilde{\sigma}_x^2}{\varepsilon_n^2 \gamma^2} = \frac{k_o^{sc} \sigma_x^2}{\varepsilon_n^2 \gamma} \gg 1 \qquad (63)$$

and ion focusing dominated when

$$\eta = \frac{4\gamma k_p^2}{3\gamma'^2} \gg 1 \ . \tag{64}$$

With the typical beam parameters of a plasma accelerator: 1 kA peak current, 2 μm normalized emittance, injection energy $\gamma_0 = 300$ and spot size about 3 μm, we have $\rho < 1$ and $\eta > 1$. It follows that the envelope equation (61) can be well approximated by the following reduced expression:

$$\sigma_x'' + \frac{k_p^2}{3\gamma}\sigma_x = \frac{\varepsilon_n^2}{\gamma^2 \sigma_x^3} \tag{65}$$

with $\gamma(z) = \gamma_0 + \alpha z$. Looking for a particular solution in the form $\sigma_x = \gamma^{-1/4}\sigma_0$ we obtain

$$\left(\frac{5}{16}\gamma'^2 + \frac{1}{3}\gamma k_p^2\right)\sigma_0 = \frac{\gamma \varepsilon_n^2}{\sigma_0^3} \tag{66}$$

that for $\eta > 1$ has a simple solution $\sigma_0 = \sqrt{\dfrac{\sqrt{3}\varepsilon_n}{k_p}}$ giving the matching condition of the beam with the plasma

$$\sigma_x = \gamma^{-1/4}\sigma_0 = \sqrt[4]{\frac{3}{\gamma}}\sqrt{\frac{\varepsilon_n}{k_p}} \ . \tag{67}$$

In Fig. 11 are shown the matched beam envelope given by Eq. (67) with normalized emittance of 2 μm and injection energy $\gamma = 300$ versus the plasma density. The figure also shows the evolution of the beam envelope in a 10 cm long plasma with density $10^{16}$ cm$^{-3}$, corresponding to an accelerating field of 5 GV/m (extraction energy $\gamma = 1300$) and focusing field of 60 MV/m.

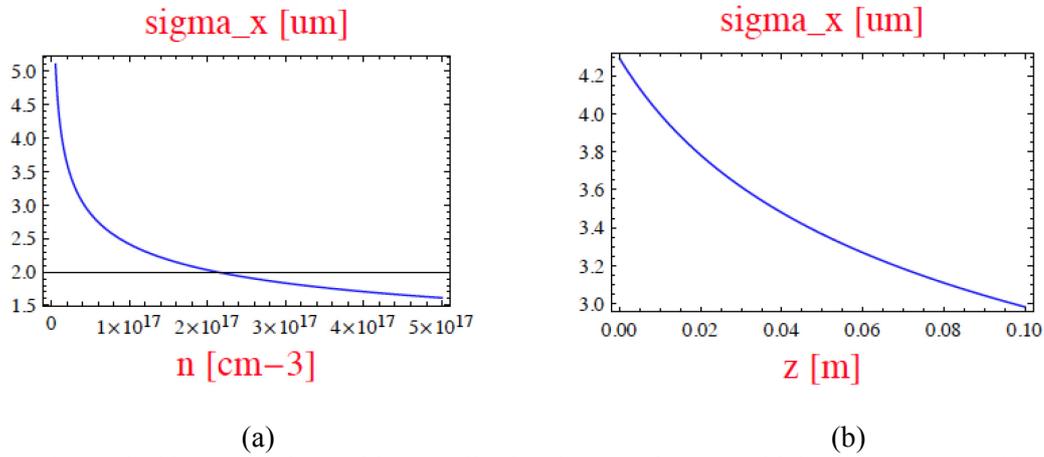

(a)           (b)

**Fig. 11:** (a) Matched beam envelope with normalized emittance of 2 μm and injection energy $\gamma = 300$ versus (b) the plasma density and the evolution of the beam envelope in a 10 cm long plasma with density $10^{16}$ cm$^{-3}$, corresponding to an accelerating field of 5 GV/m and focusing field of 60 MV/m.

Notice that the beam experiences focusing as $\gamma$ increases and the beam density increases, leading to a significant perturbation of the plasma fields. A possible solution to overcoming this effect is to taper the plasma density along the channel in order to achieve beam transport with a constant envelope.

It is an interesting exercise to see the effect of a plasma density vanishing as $n(z) = \frac{\gamma_0}{\gamma(z)} n_0$, giving $k_p^2 = \frac{e^2 n_0}{\varepsilon_0 m c^2} \frac{\gamma_0}{\gamma} = \frac{\gamma_0}{\gamma} k_{0,p}^2$. In this case the envelope equation Eq. (61) without space-charge effects becomes

$$\sigma_x'' + \frac{\gamma'}{\gamma} \sigma_x' + \frac{\gamma_0 k_{0,p}^2}{3\gamma^2} \sigma_x = \frac{\varepsilon_n^2}{\gamma^2 \sigma_x^3} \tag{68}$$

which admits a constant equilibrium solution

$$\sigma_x = \sqrt[4]{\frac{3}{\gamma_0}} \sqrt{\frac{\varepsilon_n}{k_{0,p}}}. \tag{69}$$

Figure 12 shows the plasma density along the accelerating section and the resulting equilibrium beam envelope given by Eq. (69), with the same beam parameters as those in Fig. 11

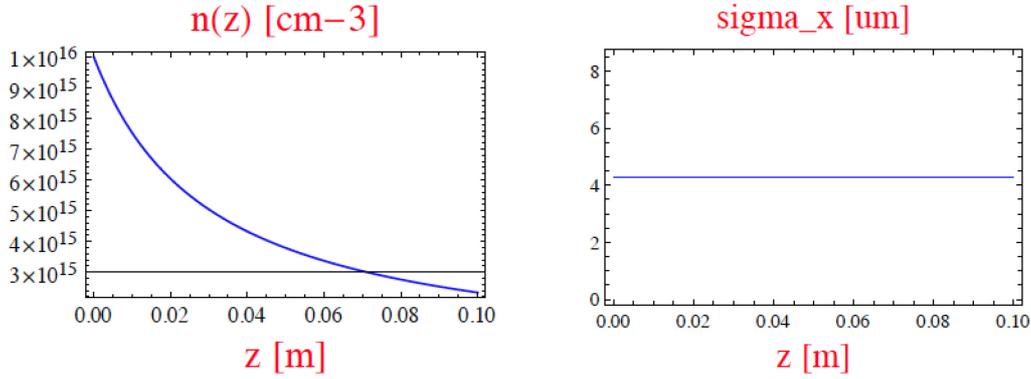

(a) (b)

**Fig. 12:** (a) Plasma density along the accelerating section; (b) the resulting equilibrium beam envelope given by Eq. (69).

On the other hand before injection in the plasma accelerator, the beam has to be focused to the matching spot given by Eq. (67) to prevent envelope oscillations that may cause emittance growth and an enhancement of betatron radiation emission. It has been proposed [21] to shape the plasma density profile in order to gently capture the beam by means of the increasing ion focusing effect. For example, by varying the plasma density as $n(z) = \frac{\gamma(z)}{\gamma_0} n_0$ at the entrance of the plasma column, the envelope equation Eq. (61) can be written as

$$\sigma_x'' + \frac{k_{0,p}^2}{3\gamma_0} \sigma_x = \frac{\varepsilon_n^2}{\gamma^2 \sigma_x^3} \tag{70}$$

where $k_p^2 = \frac{e^2 n_0}{\varepsilon_0 m c^2} \frac{\gamma}{\gamma_0} = \frac{\gamma}{\gamma_0} k_{0,p}^2$. This equation has a particular solution assuming that $\gamma''$ is negligible,

$$\sigma_x = \sqrt[4]{3\gamma_0} \sqrt{\frac{\varepsilon_n}{\gamma k_{0,p}}}, \tag{71}$$

showing that with a proper choice of the initial plasma density the beam envelope can be gently matched to the accelerating plasma channel.

For additional discussions about injection and extraction beam matching conditions see also some recent papers, Refs. [22–25].

## Acknowledgements

I wish to thank A. Cianchi, P. Muggli, J.B. Rosenzweig, A.R. Rossi and L. Serafini, for the many helpful discussions and suggestions.